\pdfoutput=1

\documentclass[12pt,a4paper]{article}

\usepackage{ifthen}
\newboolean{pdflatex}
\setboolean{pdflatex}{true}

\newboolean{articletitles}
\setboolean{articletitles}{true}

\newboolean{uprightparticles}
\setboolean{uprightparticles}{false}

\def\paperauthors{LHCb collaboration}
\def\paperasciititle{Study of nuclear effects on charm production in light-ion collisions}
\def\papertitle{Study of nuclear effects in charm production in light-ion collisions}
\def\paperkeywords{{High Energy Physics}, {LHCb}}
\def\papercopyright{\the\year\ CERN for the benefit of the LHCb collaboration}
\def\paperlicence{CC BY 4.0 licence}
\def\paperlicenceurl{https://creativecommons.org/licenses/by/4.0/}

\newif\ifEnableSectionTOCLinks
\EnableSectionTOCLinksfalse

\usepackage[top=1in, bottom=1.25in, left=1in, right=1in]{geometry}

\usepackage{microtype}
\usepackage{lineno}  
\usepackage{xspace} 
\usepackage{caption} 

\usepackage{graphicx}  
\usepackage{color}
\usepackage{colortbl}
\graphicspath{{./figs/}} 

\usepackage{amsmath} 
\usepackage{amssymb}
\usepackage{amsfonts}
\usepackage{upgreek} 

\newcommand*\patchAmsMathEnvironmentForLineno[1]{%
\expandafter\let\csname old#1\expandafter\endcsname\csname #1\endcsname
\expandafter\let\csname oldend#1\expandafter\endcsname\csname
end#1\endcsname
 \renewenvironment{#1}%
   {\linenomath\csname old#1\endcsname}%
   {\csname oldend#1\endcsname\endlinenomath}%
}
\newcommand*\patchBothAmsMathEnvironmentsForLineno[1]{%
  \patchAmsMathEnvironmentForLineno{#1}%
  \patchAmsMathEnvironmentForLineno{#1*}%
}
\AtBeginDocument{%
\patchBothAmsMathEnvironmentsForLineno{equation}%
\patchBothAmsMathEnvironmentsForLineno{align}%
\patchBothAmsMathEnvironmentsForLineno{flalign}%
\patchBothAmsMathEnvironmentsForLineno{alignat}%
\patchBothAmsMathEnvironmentsForLineno{gather}%
\patchBothAmsMathEnvironmentsForLineno{multline}%
\patchBothAmsMathEnvironmentsForLineno{eqnarray}%
}

\usepackage[pdftex,
            pdfauthor={\paperauthors},
            pdftitle={\paperasciititle},
            pdfkeywords={\paperkeywords}]{hyperref}
\usepackage{hyperxmp}
\hypersetup{
    pdfcopyright={Copyright (C) \papercopyright},
    pdflicenseurl={\paperlicenceurl}
}

\usepackage[colorinlistoftodos,textsize=scriptsize]{todonotes}

\usepackage[bottom,flushmargin,hang,multiple]{footmisc}

\usepackage[all]{hypcap} 

\usepackage{xspace} 
\usepackage{upgreek}

\def\MagUp {\mbox{\em Mag\kern -0.05em Up}\xspace}

\ifthenelse{\boolean{uprightparticles}}%
{

 \def\Pmu         {\ensuremath{\upmu}\xspace}

 \def\Ppi         {\ensuremath{\uppi}\xspace}

 \def\Ppsi        {\ensuremath{\uppsi}\xspace}

 \def\PDelta      {\ensuremath{\Delta}\xspace}                 
 \def\PXi         {\ensuremath{\Xi}\xspace}                 
 \def\PLambda     {\ensuremath{\Lambda}\xspace}                 
 \def\PSigma      {\ensuremath{\Sigma}\xspace}                 
 \def\POmega      {\ensuremath{\Omega}\xspace}                 
 \def\PUpsilon    {\ensuremath{\Upsilon}\xspace}
 \let\oldPi\Pi
 \def\PPi         {\ensuremath{\oldPi}\xspace}

 \def\PB      {\ensuremath{\mathrm{B}}\xspace}                 
 \def\PD      {\ensuremath{\mathrm{D}}\xspace}                 
 \def\PJ      {\ensuremath{\mathrm{J}}\xspace}                 
 \def\PK      {\ensuremath{\mathrm{K}}\xspace}                 
 \def\Ps      {\ensuremath{\mathrm{s}}\xspace}

 \def\thebaroffset{0.0em}
}
{

 \def\Pmu         {\ensuremath{\mu}\xspace}

 \def\Ppi         {\ensuremath{\pi}\xspace}

 \def\Ppsi        {\ensuremath{\psi}\xspace}                 
                  
 \mathchardef\PDelta="7101
 \mathchardef\PXi="7104
 \mathchardef\PLambda="7103
 \mathchardef\PSigma="7106
 \mathchardef\POmega="710A
 \mathchardef\PUpsilon="7107
 \mathchardef\PPi="7105
 \def\PB      {\ensuremath{B}\xspace}                 
 \def\PD      {\ensuremath{D}\xspace}                 
 \def\PJ      {\ensuremath{J}\xspace}                 
 \def\PK      {\ensuremath{K}\xspace}                 
 \def\Ps      {\ensuremath{s}\xspace}

 \def\thebaroffset{0.18em}
}
\newcommand{\offsetoverline}[2][\thebaroffset]{\kern #1\overline{\kern -#1 #2}}%

\makeatletter
\ifcase \@ptsize \relax
  \newcommand{\miniscule}{\@setfontsize\miniscule{4}{5}}
\or
  \newcommand{\miniscule}{\@setfontsize\miniscule{5}{6}}
\or
  \newcommand{\miniscule}{\@setfontsize\miniscule{5}{6}}
\fi
\makeatother

\DeclareRobustCommand{\optbar}[1]{\shortstack{{\miniscule (\rule[.5ex]{1.25em}{.18mm})}
  \\ [-.7ex] $#1$}}

\def\mup        {{\ensuremath{\Pmu^+}}\xspace}
\def\mun        {{\ensuremath{\Pmu^-}}\xspace} 

\def\squark    {{\ensuremath{\Ps}}\xspace}

\def\pion   {{\ensuremath{\Ppi}}\xspace}

\def\pip    {{\ensuremath{\pion^+}}\xspace}
\def\pim    {{\ensuremath{\pion^-}}\xspace}

\def\kaon    {{\ensuremath{\PK}}\xspace}

\def\KorKbar {\kern \thebaroffset\optbar{\kern -\thebaroffset \PK}{}\xspace}

\def\Km      {{\ensuremath{\kaon^-}}\xspace}

\def\KS      {{\ensuremath{\kaon^0_{\mathrm{S}}}}\xspace}

\def\D       {{\ensuremath{\PD}}\xspace}

\def\DorDbar {\kern \thebaroffset\optbar{\kern -\thebaroffset \PD}\xspace}
\def\Dz      {{\ensuremath{\D^0}}\xspace}

\def\Dp      {{\ensuremath{\D^+}}\xspace}
\def\Dm      {{\ensuremath{\D^-}}\xspace}

\def\DpDm    {\ensuremath{\Dp {\kern -0.16em \Dm}}\xspace}

\def\Dstarp  {{\ensuremath{\D^{*+}}}\xspace}

\def\B       {{\ensuremath{\PB}}\xspace}

\def\BorBbar {\kern \thebaroffset\optbar{\kern -\thebaroffset \PB}\xspace}

\def\Bd      {{\ensuremath{\B^0}}\xspace}

\def\BdorBdbar {\kern \thebaroffset\optbar{\kern -\thebaroffset \Bd}\xspace}

\def\Bs      {{\ensuremath{\B^0_\squark}}\xspace}

\def\BsorBsbar {\kern \thebaroffset\optbar{\kern -\thebaroffset \Bs}\xspace}

\def\jpsi     {{\ensuremath{{\PJ\mskip -3mu/\mskip -2mu\Ppsi}}}\xspace}

\def\Y#1S{\ensuremath{\PUpsilon{(#1S)}}\xspace}

\def\LorLbar     {\kern \thebaroffset\optbar{\kern -\thebaroffset \PLambda}\xspace}

\def\to                 {\ensuremath{\rightarrow}\xspace}

\def\AT#1     {\ensuremath{A_{\mathrm{T}}^{#1}}\xspace}           

\def\C#1      {\ensuremath{\mathcal{C}_{#1}}\xspace}                       
\def\Cp#1     {\ensuremath{\mathcal{C}_{#1}^{'}}\xspace}                    
\def\Ceff#1   {\ensuremath{\mathcal{C}_{#1}^{\mathrm{(eff)}}}\xspace}        
\def\Cpeff#1  {\ensuremath{\mathcal{C}_{#1}^{'\mathrm{(eff)}}}\xspace}       
\def\Ope#1    {\ensuremath{\mathcal{O}_{#1}}\xspace}                       
\def\Opep#1   {\ensuremath{\mathcal{O}_{#1}^{'}}\xspace}                    

\newcommand{\aunit}[1]{\ensuremath{\text{\,#1}}}       

\newcommand{\tev}{\aunit{Te\kern -0.1em V}\xspace}
\newcommand{\gev}{\aunit{Ge\kern -0.1em V}\xspace}
\newcommand{\mev}{\aunit{Me\kern -0.1em V}\xspace}
\newcommand{\kev}{\aunit{ke\kern -0.1em V}\xspace}
\newcommand{\ev}{\aunit{e\kern -0.1em V}\xspace}
 
\newcommand{\mevc}{\ensuremath{\aunit{Me\kern -0.1em V\!/}c}\xspace}
\newcommand{\gevc}{\ensuremath{\aunit{Ge\kern -0.1em V\!/}c}\xspace}
\newcommand{\mevcc}{\ensuremath{\aunit{Me\kern -0.1em V\!/}c^2}\xspace}
\newcommand{\gevcc}{\ensuremath{\aunit{Ge\kern -0.1em V\!/}c^2}\xspace}

\def\nb {\aunit{nb}\xspace}
\def\invnb {\ensuremath{\nb^{-1}}\xspace}

\newcommand{\chisqip}{\ensuremath{\chi^2_{\text{IP}}}\xspace}

\def\gsim{{~\raise.15em\hbox{$>$}\kern-.85em
          \lower.35em\hbox{$\sim$}~}\xspace}
\def\lsim{{~\raise.15em\hbox{$<$}\kern-.85em
          \lower.35em\hbox{$\sim$}~}\xspace}

\def\sqsnn {\ensuremath{\protect\sqrt{s_{\scriptscriptstyle\text{NN}}}}\xspace}
\def\pt         {\ensuremath{p_{\mathrm{T}}}\xspace}

\def\evtgen     {\mbox{\textsc{EvtGen}}\xspace}

\def\geant      {\mbox{\textsc{Geant4}}\xspace}

\def\photos     {\mbox{\textsc{Photos}}\xspace}

\def\pythia     {\mbox{\textsc{Pythia}}\xspace}

\def\tell1  {TELL1\xspace}
\def\ukl1   {UKL1\xspace}

\newcommand{\lhcborcid}[1]{\href{https://orcid.org/#1}{\hspace*{0.1em}\raisebox{-0.45ex}{\includegraphics[width=1em]{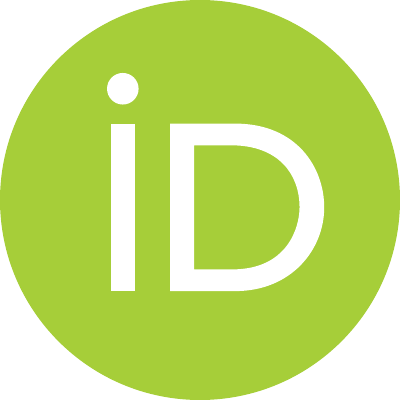}}}}

\hypersetup{
  colorlinks   = true, 
  urlcolor     = blue, 
  linkcolor    = blue, 
  citecolor    = red   
}

\ifEnableSectionTOCLinks
    \usepackage[explicit]{titlesec} 
    
    \let\oldcontentsline\contentsline
    \renewcommand

    \titleformat{\section}{\normalfont\Large\bf}{\hyperlink{tocsection.\thesection}{{\thesection} \parbox[t]{\dimexpr\textwidth-1pc}{#1}}}{1pc}{}

    \titleformat{\subsection}{\normalfont\bf}{\hyperlink{tocsubsection.\thesubsection}{{\thesubsection} \parbox[t]{\dimexpr\textwidth-1pc}{#1}}}{1pc}{}

    \titleformat{name=\section,numberless}[display]{}{}{0pt}{\normalfont\Huge\bfseries #1}
\fi

\usepackage{cite} 
\usepackage{LHCb/mciteplus}

\def\rneneoo{\ensuremath{\mathcal{R}_{\rm NeNe/OO}^{D^0}}\xspace}

\begin{document}

\renewcommand{\thefootnote}{\fnsymbol{footnote}}
\setcounter{footnote}{1}


\begin{titlepage}
\pagenumbering{roman}

\vspace*{-1.5cm}
\centerline{\large EUROPEAN ORGANIZATION FOR NUCLEAR RESEARCH (CERN)}
\vspace*{1.5cm}
\noindent
\begin{tabular*}{\linewidth}{lc@{\extracolsep{\fill}}r@{\extracolsep{0pt}}}
\ifthenelse{\boolean{pdflatex}}
{\vspace*{-1.5cm}\mbox{\!\!\!\includegraphics[width=.14\textwidth]{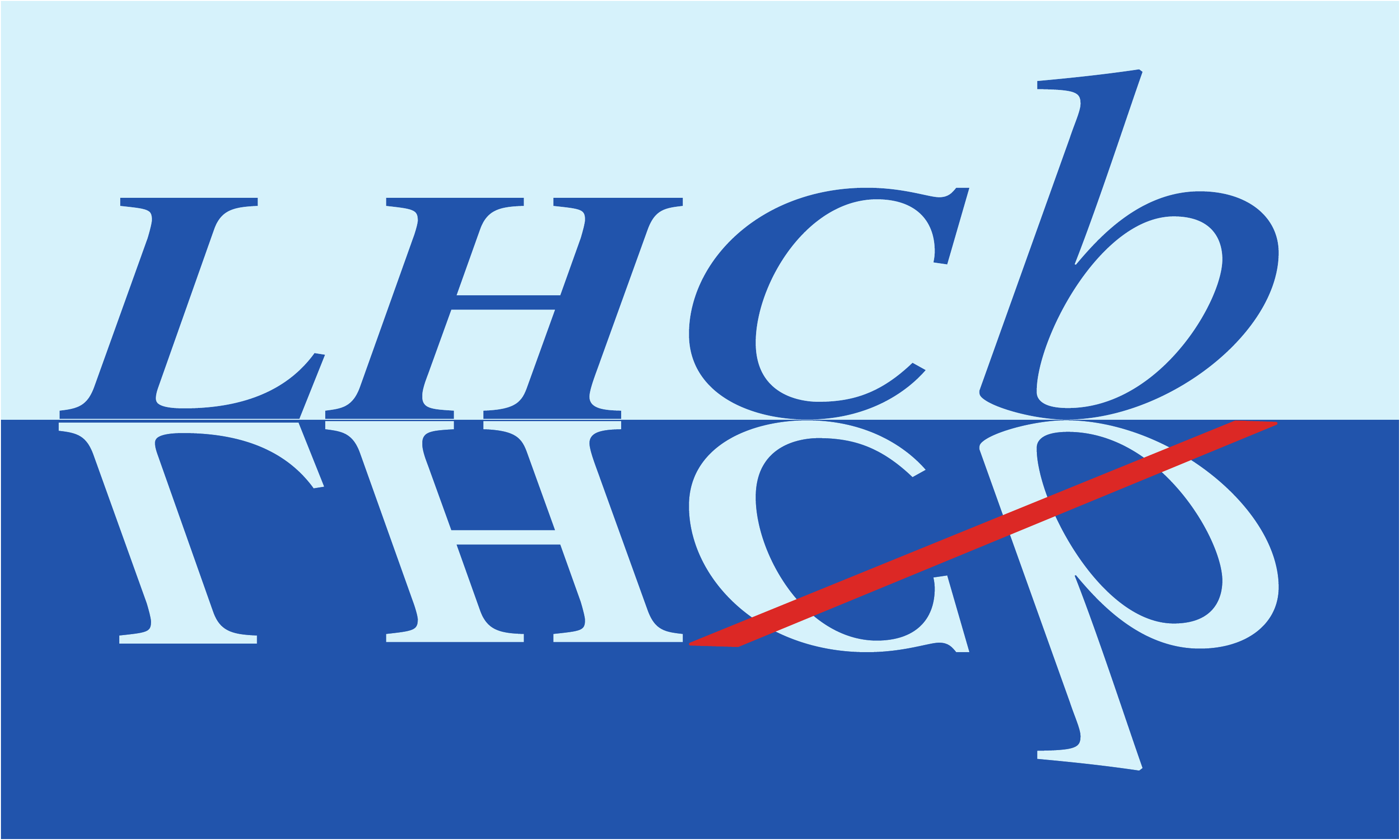}} & &}%
{\vspace*{-1.2cm}\mbox{\!\!\!\includegraphics[width=.12\textwidth]{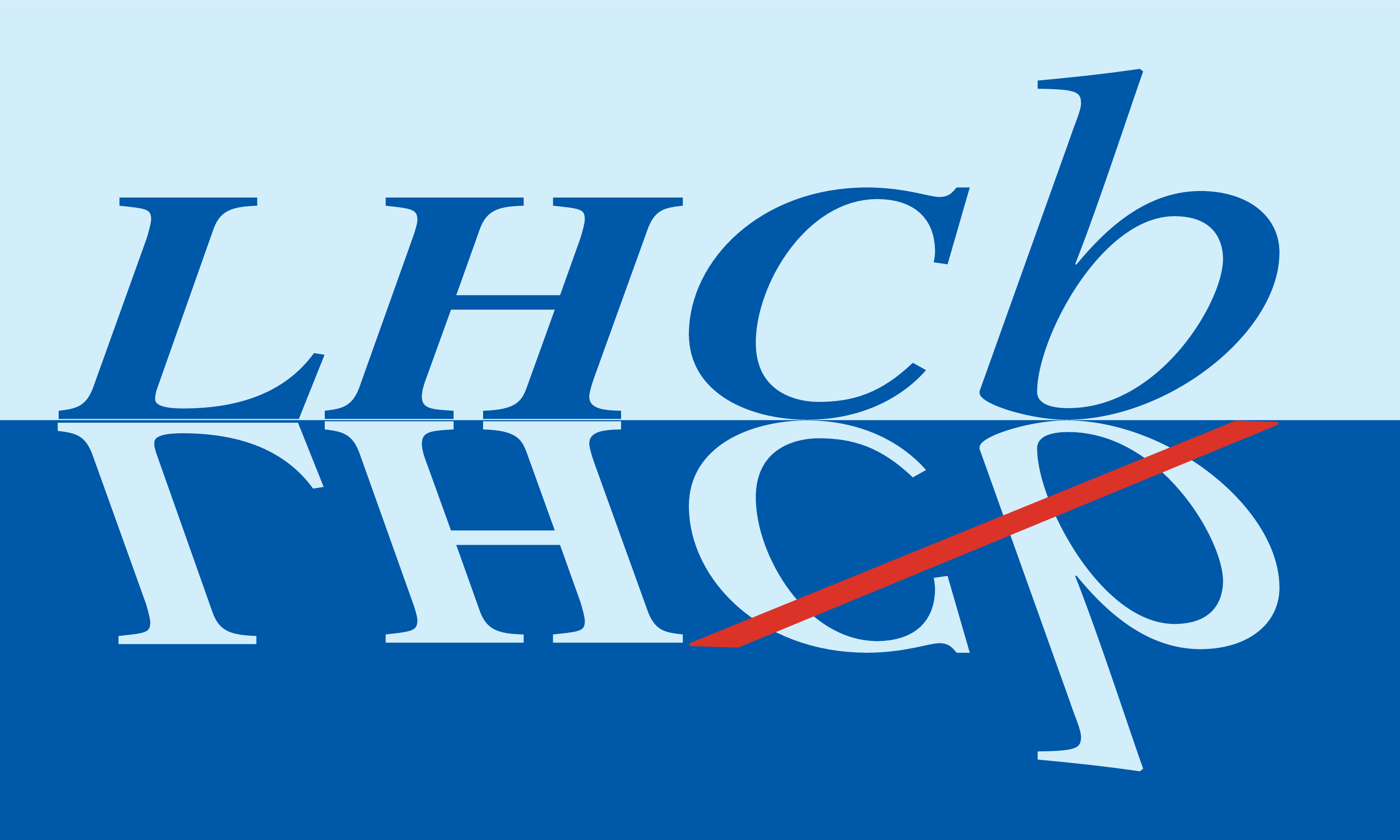}} & &}%
\\
 & & CERN-EP-2026-141 \\  
 & & LHCb-PAPER-2026-008 \\  
 & & 27 May 2026 \\ 
 & & \\
\end{tabular*}

\vspace*{4.0cm}

{\normalfont\bfseries\boldmath\huge
\begin{center}
  \papertitle 
\end{center}
}

\vspace*{2.0cm}

\begin{center}
\paperauthors\footnote{Authors are listed at the end of this paper.}
\end{center}

\vspace{\fill}

\begin{abstract}
  \noindent
  The onset of nuclear effects in light-ion collisions is studied by measuring the ratio of $D^0$ meson production between NeNe and OO collisions at a center-of-mass energy per nucleon pair of $5.36\,{\rm TeV}$ recorded by the LHCb detector.
  The $D^0$ meson yields are measured differentially in transverse momentum ($p_{\rm T}$) for \mbox{$0.5<p_{\rm T}<20\,{\rm GeV}$} in the rapidity ($y$) region $2.0<y<4.5$, and are normalized to the total number of recorded inelastic nucleus-nucleus collisions in each data sample.
  The resulting production ratio shows evidence of variation as a function of $p_{\rm T}$, which is inconsistent with predictions based on nuclear modification of nucleon structure alone.
  This measurement is consistent with the onset of quark-gluon plasma production with increasing nucleus size in high-energy nuclear collisions.
\end{abstract}

\vspace*{2.0cm}

\begin{center}
  Submitted to Phys.~Rev.~Lett.
\end{center}

\vspace{\fill}

{\footnotesize 
\centerline{\copyright~\papercopyright. \href{\paperlicenceurl}{\paperlicence}.}}
\vspace*{2mm}

\end{titlepage}


\newpage
\setcounter{page}{2}
\mbox{~}

\renewcommand{\thefootnote}{\arabic{footnote}}
\setcounter{footnote}{0}

\cleardoublepage


\pagestyle{plain} 
\setcounter{page}{1}
\pagenumbering{arabic}

The hot medium of deconfined quarks and gluons known as quark-gluon plasma (QGP) was first conclusively observed in high-energy collisions of heavy nuclei more than twenty years ago at the Relativistic Heavy Ion Collider (RHIC)~\cite{Busza:2018rrf,STAR:2005gfr,PHENIX:2004vcz,BRAHMS:2004adc,PHOBOS:2004zne}.
Since the initial discovery of QGP in heavy-ion collisions, QGP signatures such as hydrodynamic flow, enhanced production of strange hadrons, and suppression of charmonium have been observed in small collision systems such as proton-proton and proton-nucleus collisions at both RHIC and Large Hadron Collider (LHC) experiments~\cite{LHCb-PAPER-2025-003,LHCb-PAPER-2015-040,CMS:2012qk,ATLAS:2012cix,ALICE:2012eyl,PHENIX:2017xrm,PHENIX:2018lia,STAR:2022pfn,LHCb-PAPER-2023-035,LHCb-PAPER-2025-011,LHCb-PAPER-2023-021,ALICE:2016fzo}.
Evidence for parton energy loss in small collision systems, however, is notably absent~\cite{CMS:2025jbv,ATLAS:2022iyq,ALICE:2017svf}.
Explaining the unexpected observation of QGP-like signatures in small collision systems has presented a major theoretical challenge, and the origins of these effects cannot be conclusively understood with current results.

Heavy-quark production is sensitive to a wide range of QGP properties~\cite{Dong:2019byy}.
Due to their large masses, bottom and charm quarks are produced in heavy-ion collisions via high-energy parton-parton interactions before the formation of the QGP.
The heavy quarks subsequently interact with the medium throughout its evolution, losing energy through radiation and elastic collisions with QGP constituents.
Energy-loss effects are sensitive to both the microscopic interactions between the heavy quark and the QGP constituents and the path traveled by the heavy quark through the QGP.
Consequently, studies of heavy-flavor hadron production in nuclear collisions provide information about the universal transport properties of QGP and the size of the plasma produced~\cite{He:2022ywp}.
Heavy-quark production in heavy-ion collisions has been studied extensively at both RHIC and LHC experiments~\cite{ALICE:2021rxa,CMS:2024vip,ALICE:2022tji}.

In 2025, the LHC produced $^{16}$O (OO) collisions and $^{20}$Ne (NeNe) collisions for the first time at a center-of-mass energy per nucleon pair ($\sqsnn$) of \mbox{$5.36\tev$}.
The ATLAS, ALICE, and CMS collaborations have used these collisions to study the structure of $^{16}$O and $^{20}$Ne nuclei using anisotropic flow~\cite{ATLAS:2025nnt,ALICE:2025luc,CMS:2025tga}.
These light-ion collisions can also potentially create a volume of QGP large enough to produce observable parton energy loss~\cite{Huss:2020dwe,Brewer:2021kiv}. 
Evidence for parton energy loss in OO collisions was recently observed by the CMS collaboration~\cite{CMS:2025bta}.
Collisions of neon nuclei are expected to produce a slightly larger volume of QGP than those of oxygen nuclei, resulting in larger energy loss effects.
The relative magnitude of energy loss in NeNe and OO collisions is sensitive to the transport properties of the QGP and the rate of onset of QGP production with increasing system size.

This Letter presents a measurement of the ratio of $\Dz$ meson production rates in NeNe to OO collisions at \mbox{$\sqsnn=5.36\tev$} with the LHCb detector.\footnote{Charge conjugate hadrons are included implicitly throughout this Letter.}
The production ratio is defined as
\begin{equation}
    \rneneoo \equiv \frac{{\rm d}N^{\rm NeNe}_{\Dz}/{\rm d}\pt}{{\rm d}N^{\rm OO}_{\Dz}/{\rm d}\pt}\frac{N^{\rm OO}_{\rm inel}}{N^{\rm NeNe}_{\rm inel}},
    \label{eqn:rdef}
\end{equation}
where ${\rm d}N^{\rm AA}_{\Dz}/{\rm d}\pt$ is the differential $\Dz$ yield in $\rm AA$ collisions as a function of transverse momentum ($\pt$), $N^{\rm AA}_{\rm inel}$ is the number of inelastic collisions in the ${\rm AA}$ sample, and A denotes the colliding nuclei.
The  yields of $\Dz$ mesons, reconstructed in the \mbox{$\Dz\to \Km\pip$} decay, are measured differentially in $\pt$ for \mbox{$0.5<\pt<20\gev$} in the rapidity range \mbox{$2.0<y<4.5$} (natural units are used throughout this Letter).
The OO and NeNe datasets were collected in 2025 and correspond to integrated luminosities of $5.5\invnb$ and $0.51\invnb$, respectively.

The LHCb Run~3 detector~\cite{LHCb-DP-2022-002} is a single-arm forward spectrometer covering the pseudorapidity range $2<\eta<5$, designed for the study of particles containing $b$ or $c$ quarks.
It was installed during LHC long-shutdown 2, to allow effective operation at about five times the luminosity with respect to the Run 1--2 detector~\cite{LHCb-DP-2022-002,LHCb-DP-2008-001}.
The LHCb Run~3 detector keeps the general organization of subsystems of the original LHCb detector~\cite{LHCb-DP-2022-002,LHCb-DP-2008-001} though most subsystems were upgraded.
The high-precision tracking system consists of a silicon-pixel vertex detector (VELO) surrounding the interaction region, a large-area silicon-strip detector located upstream of a dipole magnet, and three stations of scintillating-fiber detectors located downstream of the magnet~\cite{LHCb-TDR-015}.
Different types of charged hadrons are distinguished using information from two ring-imaging Cherenkov (RICH) detectors~\cite{LHCb-TDR-014}.

Simulated samples are used to model the detector response to $\Dz$ decays.
In the simulation, $\Dz$ mesons are generated using \pythia~\cite{Sjostrand:2007gs} with a specific LHCb configuration and embedded into OO and NeNe events generated using EPOS-LHC~\cite{Pierog:2013ria}.
Decays of unstable particles are described by \evtgen~\cite{Lange:2001uf} and final-state radiation is generated using \photos~\cite{davidson2015photos}.
The interaction of the generated particles with the detector, and its response, are implemented using the \geant toolkit~\cite{Agostinelli:2002hh,Allison:2006ve} as described in Ref.~\cite{LHCb-PROC-2011-006}.

The ratio $\rneneoo$ can be expressed in terms of experimental quantities,
\begin{equation}
  \rneneoo=\frac{Y^{\rm NeNe}_{D^0}}{Y^{\rm OO}_{D^0}}\times\frac{\epsilon_{\rm OO}}{\epsilon_{\rm NeNe}}\times\frac{N_{\rm inel}^{\rm OO}}{N_{\rm inel}^{\rm NeNe}},
  \label{eqn:observable}
\end{equation}
where $Y^{\rm AA}_{D^0}$ and $\epsilon_{\rm AA}$ are the $\Dz$ yields and detection efficiencies in each $\pt$ region, respectively.
All NeNe and OO collision events must have at least one reconstructed track in the VELO.
Candidate $\Dz$ mesons are reconstructed using the decay \mbox{$\Dz\to K^-\pi^+$}.
The $\Km$ and $\pip$ candidates consist of good-quality tracks that are inconsistent with originating from the collision vertex. 
They must have $\pt>400\mev$, $p>5\gev$, and $2<\eta<5$, and they must be identified as kaons and pions, respectively, using information from the RICH system.
The final-state tracks must form a good-quality $\Dz$ decay vertex, and the $\Dz$ candidate's momentum must point from the collision vertex to the decay vertex.

\begin{figure}
  \centering
  \includegraphics[width=\textwidth]{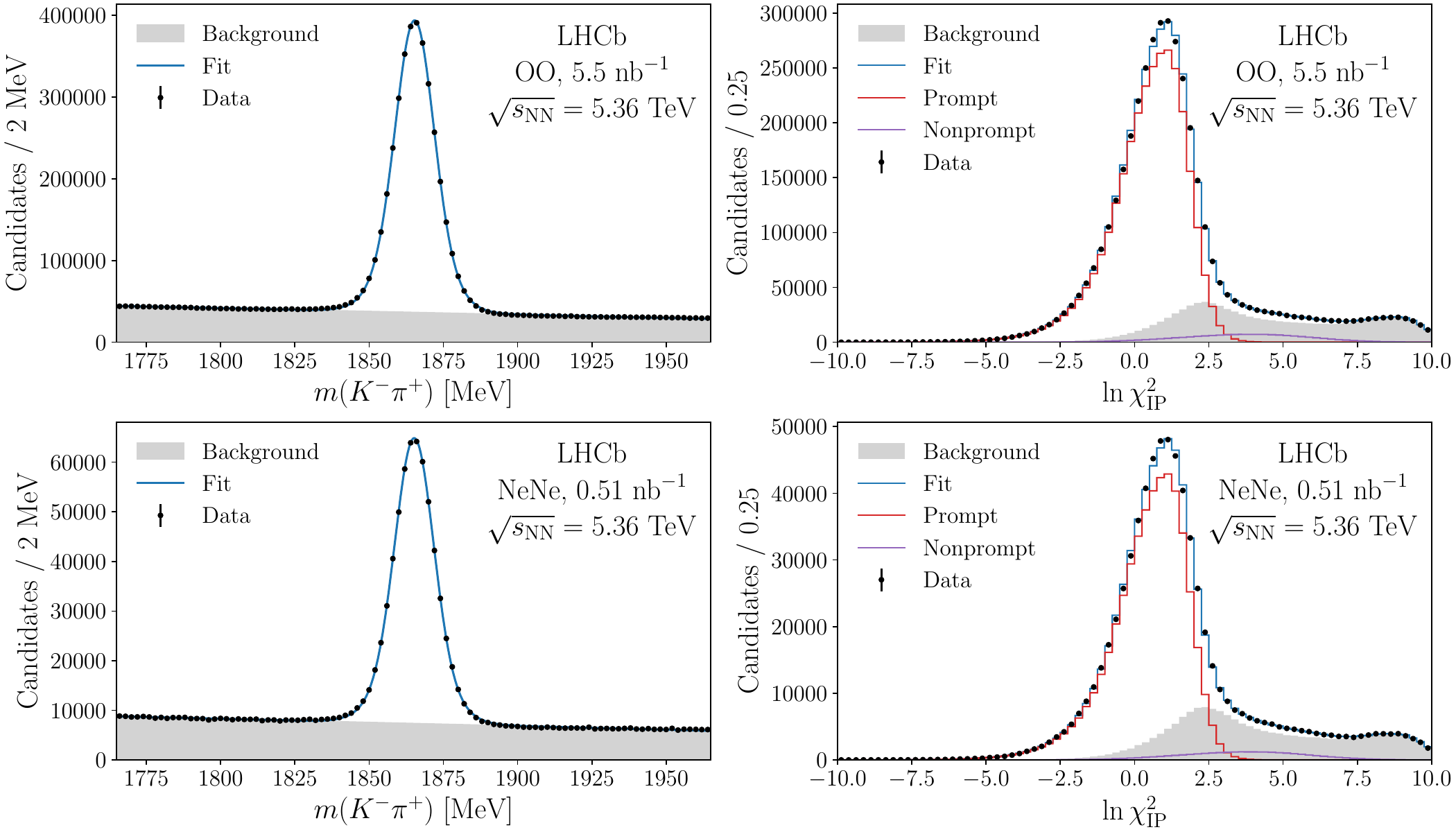}
  \caption{Distributions for reconstructed and selected $\D^0\to K^-\pi^+$ candidates: (left) $K^-\pi^+$ mass distributions and (right) $\ln\chi^2_{\rm IP}$ distributions in the $m(K^-\pi^+)$ signal window for (top) OO and (bottom) NeNe collisions.
  The sum of the fits in each $p_{\rm T}$ interval are also shown.}
  \label{fig:fits}
\end{figure}

The selected $\Dz$ samples include prompt $\Dz$ signal produced directly from the collision, nonprompt $\Dz$ mesons produced in $b$-hadron decays, and combinatorial background.
The combinatorial background contribution is determined using a binned maximum-likelihood fit to the $\Dz$ candidate mass distribution $m(\Km\pip)$~\cite{py:iminuit}.
The $\Dz$ signal is modeled using the sum of two Gaussian functions with a common mean, and the combinatorial background is modeled using a linear function.
The prompt $\Dz$ signal yield is then determined using a fit to the $\Dz$ candidate $\ln\chisqip$ distribution, where $\chisqip$ is defined as the difference in $\chi^2$ of the collision vertex fit performed with and without the $\Dz$ candidate.
Since $\Dz$ mesons from $b$-hadron decays do not originate from the collision vertex, they have larger average $\chisqip$ values than prompt $\Dz$ mesons.
The $\ln\chisqip$ fit is performed for candidates in the signal mass region, defined as a $\pm25\mev$ window around the known $\Dz$ mass~\cite{PDG2024}.
The combinatorial background yield in the signal region is fixed using the results of the $m(\Km\pip)$ fit.
Candidates outside the signal region are used to construct a combinatorial background template.
The prompt signal model is a bifurcated Gaussian function with exponential tails and is described in Ref.~\cite{LHCb-PAPER-2015-041}. The nonprompt $\Dz$ contribution is modeled using a bifurcated Gaussian function.
The prompt and nonprompt model parameters are constrained using fits to simulated prompt $\Dz$ mesons and $\Dz$ mesons from semileptonic $b$-hadron decays, respectively.
The $m(\Km\pip)$ and $\ln\chisqip$ fitting procedure is repeated in each $\pt$ region.
The sums of the fits over all $\pt$ regions are shown in Fig.~\ref{fig:fits}.

The $\Dz$ detection efficiency in Eq.~\ref{eqn:observable} includes contributions from reconstruction, charged-particle identification (PID), and selection efficiencies.
These efficiencies are similar in OO and NeNe collisions, but the difference in charged-particle multiplicities between the two systems leads to small differences, which are measured in data.
Due to the high granularity of the VELO detector, the VELO track reconstruction efficiency does not vary significantly within the charged-particle multiplicities produced in OO and NeNe collisions.
As a result, only the efficiency of reconstructing a charged particle traversing the full detector, given a reconstructed VELO track segment, is corrected.
The charged-particle reconstruction efficiency is measured using $\KS\to\pip\pim$ decays.
The $\KS$ travels a macroscopic distance before decaying, creating a displaced decay vertex.
One of the final-state pions is reconstructed as a track with signals in both the VELO and scintillating-fiber detectors (long track).
Because this track includes signals both upstream and downstream of the dipole magnet, its momentum is measured precisely.
The other ``probe'' pion is reconstructed using only signals from the VELO.
The probe pion momentum is calculated by requiring that the total $\KS$ momentum points along its reconstructed flight direction.
The flight direction is defined by the vector pointing from the collision vertex to the $\KS$ decay vertex.
The track reconstruction efficiency is then the fraction of reconstructed $\KS$ mesons with probe tracks that have a corresponding reconstructed long track.
This efficiency is measured as a function of charged particle kinematics and the total number of activated VELO pixel clusters in the collision event ($N_{\rm VP}$).

The PID efficiency is measured using the \mbox{$\Dz\to\Km\pip$} decay products in \mbox{$\Dstarp\to\Dz\pip$} decays.
The reconstructed $\Dz$-meson candidate and its decay products must satisfy the same kinematic and reconstruction quality requirements as those used for determining the total $\Dz$ yields, but no PID criteria are required.
The accompanying pion is required to have $p>1\gev$, $\pt>200\mev$, $2<\eta<5$, and must be consistent with originating from the collision vertex.
The PID efficiency is measured separately for kaons and pions and corresponds to the fraction of candidates for which the relevant hadron is correctly identified.
This efficiency is also measured as a function of the hadron kinematics and $N_{\rm VP}$.

The charged-particle reconstruction and identification efficiencies are used to estimate the total $\Dz$ reconstruction and PID efficiencies.
The $\Dz$ efficiency is calculated by convolving the measured reconstruction and PID efficiency maps with the kinematic and multiplicity distributions from simulated $\Dz$ decays.
In order to accurately describe the OO and NeNe event multiplicities, the $N_{\rm VP}$ distributions in the simulated samples are corrected to match those in data.
The combined reconstruction and PID efficiency ratio varies between about 1.03 and 1.04 in all $\pt$ intervals, except for the highest-$\pt$ interval where the correction is \mbox{$1.07\pm0.03$}.

Because quantities related to track and decay vertex quality do not depend on multiplicity, the efficiency of requirements based on these quantities will cancel in $\rneneoo$.
However, due to their higher multiplicities NeNe collision vertices are reconstructed with better resolution than OO collision vertices, leading to better IP resolution for $\Dz$ decay products.
Given that $\chisqip$ is inversely proportional to the square of the IP resolution, this difference results in different efficiencies of the track $\chisqip$ requirement in the two samples.
The magnitude of this effect is estimated by measuring the impact parameter (IP) resolution for charged particles in NeNe and OO collisions using \mbox{$\jpsi\to\mup\mun$} decays.
The IP resolutions in NeNe and OO collisions are found to be compatible within an uncertainty of about $0.8\%$, and the resulting $\rneneoo$ uncertainty ranges between 0.4 and 1.0\%.
The difference in collision vertex reconstruction performance also affects the efficiency of the $\Dz$ momentum pointing requirement.
This efficiency difference is estimated by measuring the $\Dz$ yield with and without the pointing requirement.
The measured efficiency difference has a maximum of about $0.8\%$ in the lowest $\pt$ interval and is less than $0.5\%$ in most other intervals.

The uncertainties are summarized in Table~\ref{tab:uncertainties} and fall into three categories: uncorrelated between $\pt$ intervals (uncorrelated), correlated and $\pt$-dependent (correlated), and correlated and constant in $\pt$ (global).
The largest uncertainty in most $\pt$ intervals is a global systematic uncertainty from the determination of $N_{\rm inel}^{\rm OO}/N_{\rm inel}^{\rm NeNe}$.
The $N_{\rm inel}$ ratio is estimated using the ratio of visible collisions in the OO and NeNe samples.
The uncertainty is estimated by varying the criteria used to define a visible collision.
This uncertainty accounts for possible differences in the probabilities of inelastic NeNe and OO collisions producing visible activity in the detector.
The baseline definition requires a reconstructed collision vertex in the interaction region, while the alternative requires two VELO tracks.
The relative difference in $N_{\rm inel}^{\rm OO}/N_{\rm inel}^{\rm NeNe}$ between these two definitions is $4.1\%$.

\begin{table}[!tb]
  \centering
  \caption{Relative sources of uncertainties in percent.
  The ranges correspond to the minimum and maximum values of the associated uncertainties across all $p_{\rm T}$ intervals.}
  \begin{tabular}{l c}
    \hline
    Source & Uncertainty [\%]\\
    \hline
    Uncorrelated&\\
    \hline
    Statistical uncertainty & 0.5--5.7\\
    Simulation sample size & 0.3--5.2\\
    Pointing efficiency & 0.2--0.8\\
    \hline
    Correlated&\\
    \hline
    Fit model & 0.1--0.5\\
    Rec. and PID efficiency & 0.2--1.7\\
    IP resolution & 0.4--1.0\\
    \hline
    Global&\\
    \hline
    Normalization & 4.1\\
    \hline
  \end{tabular}
  \label{tab:uncertainties}
\end{table}
Uncorrelated uncertainties arise from the finite size of the experimental and simulated datasets, as well as the statistical uncertainty associated with the efficiency of the $\Dz$ pointing requirement.
The largest correlated uncertainty in most $\pt$ intervals arises from the uncertainty in the IP resolution difference between NeNe and OO collisions.
This effect is estimated by scaling the minimum $\chisqip$ requirement in one sample at a time.
The scaling factor is given by the square of the measured IP resolution ratio.
This ratio is varied by its uncertainty, and the deviation from the baseline result is taken as a systematic uncertainty.
An additional correlated systematic uncertainty arises from the choice of fit model for prompt $\Dz$ yield determination.
The measurement is repeated using alternative fit models, and the largest absolute deviation from the baseline in each $\pt$ interval is taken as the systematic uncertainty, which is fully correlated across $\pt$ intervals.

Correlated uncertainties also arise from the limited sample sizes of the tracking and PID efficiency calibration samples.
These are estimated by repeating the measurement using bootstrapped calibration samples and taking the standard deviation of the resulting variations as the systematic uncertainty.
The PID efficiency is also affected by the choice of kinematic intervals used for the calibration, as the PID efficiency varies rapidly at low momentum near the threshold for producing Cherenkov radiation in the RICH detector.
The measurement is repeated with finer momentum intervals, and the difference between this and the baseline measurement is taken as a systematic uncertainty.
Kinematic differences between data and simulation can also result in an incorrect efficiency determination.
This effect is studied by calculating the efficiency using the kinematic and $N_{\rm VP}$ distributions of reconstructed $\Dz$ candidates in data.
The difference between this and the baseline result is taken as a systematic uncertainty.
This uncertainty is subdominant everywhere except the two lowest $\pt$ intervals, where it is the largest correlated systematic uncertainty.
All reconstruction and PID efficiency uncertainties are combined in Table~\ref{tab:uncertainties}.
A covariance matrix is calculated for each uncertainty source, and the total covariance matrix is the sum of these contributions.

\begin{figure}[!tb]
  \centering
  \includegraphics[width=0.48\textwidth]{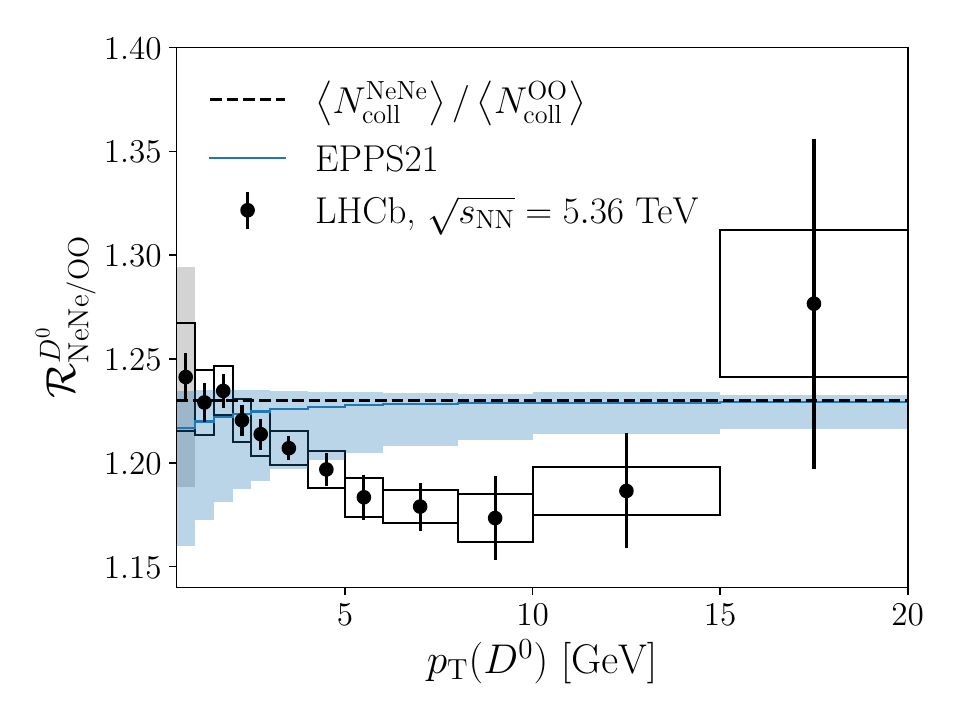}
  \includegraphics[width=0.48\textwidth]{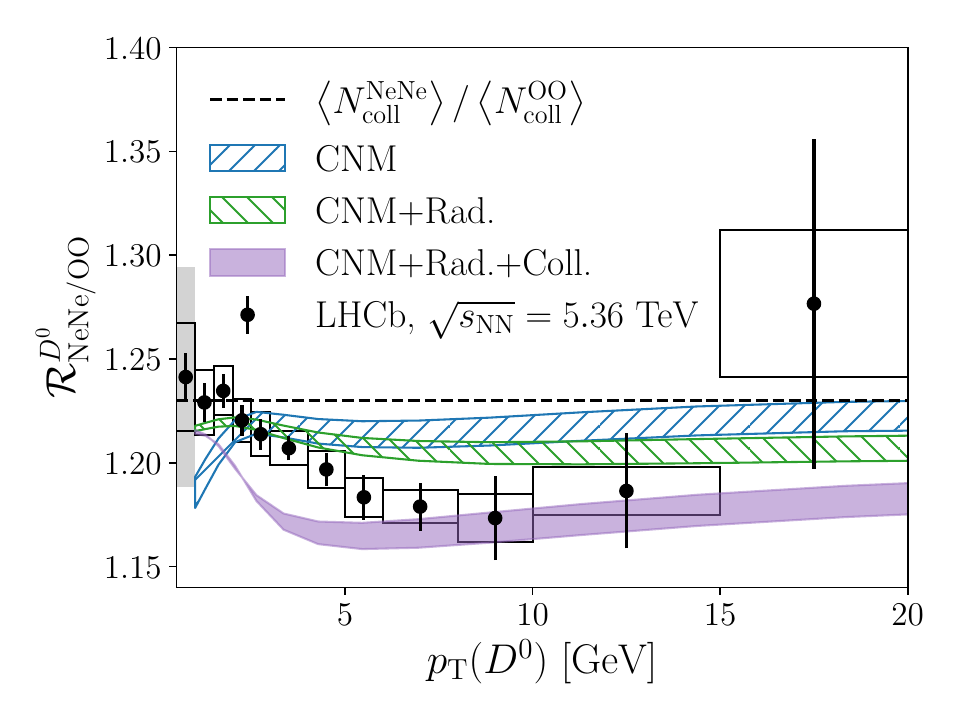}
  \caption{
    Production ratio $\mathcal{R}_{\rm NeNe/OO}^{D^0}$ in regions of $D^0$ transverse momentum.
    Uncorrelated uncertainties are shown as vertical bars, and correlated uncertainties are shown as boxes. The horizontal widths of the boxes correspond to the $p_{\rm T}$ interval.
    The global normalization uncertainty is shown as a gray band centered on the lowest $p_{\rm T}$ data point.
    The left panel shows a calculation using the EPPS21 nPDF set.
    The blue band shows the nPDF uncertainties for a $68\%$ confidence interval.
    This calculation is scaled by the ratio of the average number of binary nucleon-nucleon collisions per nuclear collision, which is shown as a black dashed line~\cite{Loizides:2025ule}.
    No uncertainty from this scaling factor is shown, although this normalization uncertainty is expected to be large relative to both the measurement and nPDF uncertainties.
    The right panel shows calculations that include cold nuclear matter effects (CNM), as well as radiative (Rad.) and collisional (Coll.) energy loss in the QGP~\cite{Ke:2022gkq}.
    }
    \label{fig:results}
\end{figure}

The fully corrected $\rneneoo$ ratio is shown in Fig.~\ref{fig:results}.
The data are compared to a next-to-leading order perturbative quantum chromodynamics (pQCD) calculation of the $\Dz$ cross section ratio using the EPPS21 nuclear parton distribution function (nPDF) set~\cite{Eskola:2021nhw,Helenius:2018uul}. 
This calculation is scaled by the ratio of the average number of binary nucleon-nucleon collisions in a single nuclear collision ($\left<N_{\rm coll}^{\rm OO}\right>/\left<N_{\rm coll}^{\rm NeNe}\right>$)~\cite{Loizides:2025ule}.
The pQCD calculation accounts for differences in the partonic structure between nucleons in neon and oxygen nuclei.
This calculation does not describe the shape of the measured ratio, indicating the presence of nuclear effects beyond the modification of nuclear parton densities.
The data are also compared to calculations that include cold nuclear matter (CNM) effects, such as parton distribution modification, as well as radiative and collisional energy loss in the QGP~\cite{Ke:2022gkq}.
Only the calculation including both collisional and radiative energy loss in the QGP accurately reproduces the $\pt$ dependence of the measured ratio.
Consequently, these data suggest the presence of QGP-like effects in charm production in NeNe collisions. 
These QGP-like effects are stronger in NeNe collisions than in OO collisions, consistent with a gradual increase in QGP production with increasing collision size.

In conclusion, the $\Dz$-meson production ratio $\rneneoo$ is measured at forward rapidity.
This is the first study of charm production in light-ion collisions at the LHC.
The large light-ion datasets collected in 2025 and the excellent performance of the LHCb detector allow for precise measurements of this ratio over a wide kinematic range.
The resulting measurement shows evidence for the gradual increase of QGP-like effects with collision size in light-ion collisions.
Furthermore, the high precision of this measurement can provide constraints on heavy quark diffusion and energy loss in the produced medium.

\section*{Acknowledgements}
%
%
\noindent We express our gratitude to our colleagues in the CERN
accelerator departments for the excellent performance of the LHC. We
thank the technical and administrative staff at the LHCb
institutes.
We acknowledge support from CERN and from the national agencies:
ARC (Australia);
CAPES, CNPq, FAPERJ and FINEP (Brazil); 
MOST and NSFC (China); 
CNRS/IN2P3 and CEA (France);  
BMFTR, DFG and MPG (Germany);
INFN (Italy); 
NWO (Netherlands); 
MNiSW and NCN (Poland); 
MEC/IFA (Romania); 
MICIU and AEI (Spain);
SNSF and SER (Switzerland); 
NASU (Ukraine); 
STFC (United Kingdom); 
DOE NP and NSF (USA).
We acknowledge the computing resources that are provided by ARDC (Australia), 
CBPF (Brazil),
CERN, 
IHEP and LZU (China),
IN2P3 (France), 
KIT and DESY (Germany), 
INFN (Italy), 
SURF (Netherlands),
Polish WLCG (Poland),
IFIN-HH (Romania), 
PIC (Spain), CSCS (Switzerland), 
GridPP (United Kingdom),
and NSF (USA).  
We are indebted to the communities behind the multiple open-source
software packages on which we depend.
Individual groups or members have received support from
RTP (Australia), 
FWO Odysseus grant G0ASD25N (Belgium), 
Key Research Program of Frontier Sciences of CAS, CAS PIFI, CAS CCEPP (China); 
Minciencias (Colombia);
EPLANET, Marie Sk\l{}odowska-Curie Actions, ERC and NextGenerationEU (European Union);
A*MIDEX, ANR, IPhU and Labex P2IO, and R\'{e}gion Auvergne-Rh\^{o}ne-Alpes (France);
Alexander-von-Humboldt Foundation (Germany);
ICSC (Italy); 
Severo Ochoa and Mar\'ia de Maeztu Units of Excellence, GVA, XuntaGal, GENCAT, InTalent-Inditex and Prog.~Atracci\'on Talento CM (Spain);
the Leverhulme Trust, the Royal Society and UKRI (United Kingdom).

\addcontentsline{toc}{section}{References}
\bibliographystyle{LHCb/LHCb}
\bibliography{main,LHCb/standard,LHCb/LHCb-PAPER,LHCb/LHCb-CONF,LHCb/LHCb-DP,LHCb/LHCb-TDR}

\newpage

\centerline
{\large\bf LHCb collaboration}
\begin
{flushleft}
\small
R.~Aaij$^{38}$\lhcborcid{0000-0003-0533-1952},
M.~Abdelfatah$^{69}$,
A.S.W.~Abdelmotteleb$^{57}$\lhcborcid{0000-0001-7905-0542},
C.~Abellan~Beteta$^{51}$\lhcborcid{0009-0009-0869-6798},
F.~Abudin\'en$^{59}$\lhcborcid{0000-0002-6737-3528},
T.~Ackernley$^{61}$\lhcborcid{0000-0002-5951-3498},
A.A.~Adefisoye$^{69}$\lhcborcid{0000-0003-2448-1550},
B.~Adeva$^{47}$\lhcborcid{0000-0001-9756-3712},
M.~Adinolfi$^{55}$\lhcborcid{0000-0002-1326-1264},
P.~Adlarson$^{87,42}$\lhcborcid{0000-0001-6280-3851}
% ...

{\footnotesize \it

% ...
$^{38}$Nikhef National Institute for Subatomic Physics, Amsterdam, Netherlands\\
% ...
$^{42}$National Center for Nuclear Research (NCBJ), Warsaw, Poland\\
% ...
$^{47}$Instituto Galego de F{\'\i}sica de Altas Enerx{\'\i}as (IGFAE), Universidade de Santiago de Compostela, Santiago de Compostela, Spain\\
% ...
$^{51}$Physik-Institut, Universit{\"a}t Z{\"u}rich, Z{\"u}rich, Switzerland\\
% ...
$^{55}$H.H. Wills Physics Laboratory, University of Bristol, Bristol, United Kingdom\\
% ...
$^{57}$Department of Physics, University of Warwick, Coventry, United Kingdom\\
% ...
$^{59}$School of Physics and Astronomy, University of Edinburgh, Edinburgh, United Kingdom\\
$^{60}$School of Physics and Astronomy, University of Glasgow, Glasgow, United Kingdom\\
$^{61}$Oliver Lodge Laboratory, University of Liverpool, Liverpool, United Kingdom\\
% ...
$^{69}$Syracuse University, Syracuse, NY, United States\\
% ...
$^{87}$Department of Physics and Astronomy, Uppsala University, Uppsala, Sweden, associated to $^{60}$\\
% ...
}
\end{flushleft}

\end{document}